# Assessment of active dopants and p-n junction abruptness using in-situ biased 4D-STEM

Bruno C. da Silva*[1], Zahra S. Momtaz[1], Eva Monroy[2], Hanako Okuno[3], Jean-Luc Rouviere[3], David Cooper[4] and Martien I. den-Hertog[1]

[1]*Univ. Grenoble Alpes, CNRS-Institut Néel, F-38000 Grenoble, France*

[2]*Univ. Grenoble Alpes, CEA, Grenoble INP, IRIG, PHELIQS, F-38000 Grenoble, France*

[3]*Univ. Grenoble Alpes, CEA, IRIG, MEM, LEMMA, F-38000 Grenoble, France*

[4]*Univ. Grenoble Alpes, CEA-LETI, F-38000 Grenoble, France*

*[bruno-cesar.da-silva@neel.cnrs.fr](mailto:bruno-cesar.da-silva@neel.cnrs.fr)*

**Abstract**

A key issue in the development of high-performance semiconductor devices is the ability to properly measure active dopants at the nanometer scale. 4D scanning transmission electron microscopy and off-axis electron holography have opened up the possibility of studying the electrostatic properties of a p-n junction with nm-scale spatial resolution. The complete description of a p-n junction must take into account the precise evolution of the concentration of dopants around the junction, since the sharpness of the dopant transition directly influences the built-in potential and the maximum electric field. Here, a contacted silicon p-n junction is studied through in-situ biased 4D-STEM. Measurements of electric field, built-in voltage, depletion region width and charge density in the space charge region are combined with analytical equations as well as finite-element simulations in order to evaluate the quality of the junction interface. The nominally-symmetric, highly doped ($N_A = N_D = 9 \times 10^{18}\ cm^{-3}$) junction presents an electric field and built-in voltage much lower than expected for an abrupt junction. These experimental results are consistent with electron holography data. All measured junction parameters are compatible with the presence of an intermediate region with a graded profile of the dopants at the p-n interface. This hypothesis is also consistent with the evolution of the electric field with bias. These results demonstrate that in-situ biased 4D-STEM enables a better understanding of the electrical properties of semiconductor p-n junctions with nm-scale resolution.

Increasingly stringent performance demands for semiconductor devices have led to downsizing device regions to the nanoscale. In such nanodevices, the functionality is often determined by the concentration and profile of electrically active dopants.[1–4] Therefore, quantitative methods to measure the electrically active dopant density and profile with nm-scale resolution and high sensitivity are required for the successful engineering of device performance.

Profiling the dopant concentration in a semiconductor can be done by Secondary Ion Mass Spectrometry (SIMS)[5] or Atom Probe Tomography (APT).[6] However, these techniques are destructive and only sensitive to the chemical nature of dopants, and not to their electrical activity. Another view is given by electrical transport measurements, such as Hall effect and capacitance-voltage (CV) measurements.[4,7,8] Standard Hall effect can measure the free carrier density and mobility, but not the dopant profile. In contrast, CV characterization offers a good approach to the profile of active dopants, and measures the depletion region width. However, it is limited in 'depth', and it provides an averaged view of the n and p regions, unless the junction is extremely assymetric.[4,9,10] The electrochemical version of CV measurements (ECV) is not 'depth limited', but it is destructive.

Techniques based on transmission electron microscopy (TEM) can provide access to the profile of active dopants[11], defects or polarization fields[12], with nanometer spatial resolution. Off-axis electron holography, for example, has successfully been used to measure the built-in potential and estimate the active dopant concentration with spatial resolution down to 1 nm.[13–16] This technique can image built-in potential across lamella samples prepared from bulk p-n junctions or nanomaterials such as nanowires[15,17], and can be used combined with in-situ biasing[13,17]. Holography requires a vacuum region close to the sample for the reference wave, and this can be difficult to combine with a reliable electrical contact in the case of nanomaterials[18]. Approaches based on Scanning (Transmission) Electron Microscopy (SEM or STEM), such as Electron-Beam Induced Current (EBIC)[19,20] and Differential Phase Contrast (DPC)[21,22], are also able to study the electrostatics of a p-n junction. Unfortunately, quantification from DPC with a segmented detector is not straightforward and depends on the so called hard shift of the disk to provide reliable results[23], while in EBIC the electric field is mapped qualitatively and is convoluted with the carrier diffusion length. In both cases, quantitative information can be challenging or impossible to extract.

Advances in detector technology have led to the development of four-dimensional (4D) STEM, where the conventional segmented annular dark field detectors are replaced by high-speed pixelated devices, allowing the recording of the convergent beam electron diffraction (CBED) pattern at each scan point of the electron beam, resulting in an electron diffraction map, which is a 4D data map (2 dimensions for scanning directions and 2 directions in the diffraction patterns). In particular, the implementation of momentum-resolved 4D-STEM has allowed to measure internal electric fields with ultrahigh spatial resolution.[24–26] This has been demonstrated also for long range built-in electric fields in GaAs,[26] but the low signal-to-noise ratio (SNR) reported can be a drawback to recover the charge density through the derivation of the measured electric field. Recently, we have demonstrated that nano-beam STEM mode can provide reliable, electric field maps in a p-n junction even with low electron dose.[27]

In the present study, we report in-situ biased 4D-STEM experiments, using a high-speed pixelated direct electron detector with Medipix3 technology, performed on a Si p-n junction with electrical contacts. The high speed acquisition enables to acquire large field of view maps at different biases without sample drift. This approach leads to reliable measurements of the charge density, electric field and electrostatic potential as a function of bias, with a larger field of view. The interpretation of the results does not require advanced modeling of the junction and the measurement do not need the presence of a vacuum region close to the specimen. Measurements are compared to finite element simulations, as well as SIMS and electron holography measurements performed in the same wafer.

The sample under study is a symmetrically doped silicon p-n junction grown by reduced-pressure chemical vapor deposition. Dopants were phosphorus for the *n*-type region and boron for the *p*-type segment, both at a concentration of $9 \times 10^{18}\ cm^{-3}$. The lamella specimen was mounted in a Protochips Aduro 500 system and prepared using a FEI Strata dual beam focused ion beam (FIB) tool. The specimen was welded by FIB-assisted metal (W) deposition across a biasing chip, and isolation cuts were made so that bias is applied across the p-n junction. The specimen was prepared using optimized protocols to reduce gallium implantation.[28] The lamella was 350 nm thick, with an electrically inactive region thickness $t_{dead} = (60 \pm 15)$ nm at the surfaces.[19] 4D-STEM was performed in a FEI Titan Ultimate aberration-corrected (S)TEM microscope operated at 200 kV and equipped with a fast pixelated (256 × 256 pixels) Medipix3-based Merlin camera. Measurements were performed in nano-beam STEM mode, using a nominal semi-convergence angle of 1.09 mrad and a camera length of 2.3 m. With these settings, there are no overlapping disks in the CBED patterns, and the pixelated detector records the transmitted beam. The probe diameter was estimated from a TEM image (Figure S1a); its full width at half maximum is about 3.6 nm. In reciprocal space, the angular resolution (the pixel size in the pixelated detector) was 11.0 μrad. Electron diffraction patterns of 178 × 105 pixels were acquired in the single-pixel mode of the Merlin detector, with a threshold $t_0 = 80$ kV. A representative diffraction pattern is shown in Figure S1b. Experiments were performed using a beam current of 100 pA, a step size of 5 nm and a dwell time for each diffraction pattern of 5 ms. The sample was tilted off-axis to minimize the diffraction contrast while keeping the p-n junction perpendicular to the direction of the electron beam. We noticed that the DPC signal is clearer in off-axis conditions, therefore live DPC on a 4-quadrant DPC detector was used to establish a crystal orientation with reduced diffraction contrast. Biased measurements were carried out by recording the 4D dataset while bias was applied on the sample using an external Keithley source. Figure 1 shows a schematic description of the momentum-resolved 4D-STEM experiments on the Si lamella containing a p-n junction. Reverse (forward) bias is used to increase (decrease) the electric field at the junction. Data analysis of the 4D datasets was carried out with python scripts using basic open-source libraries (e.g. HyperSpy).[29] The procedure to extract the projected electric field maps is described elsewhere.[22,26,27] The spatial resolution of the space charge region is fundamentally limited by the Debye length $L_D$, which depends on the charge density in the border of the space charge region.[1,30,31] Therefore, the spatial resolution of the electric field maps can be limited by either, the probe size or $L_D$, depending on the charge density. The p-n junction was previously characterized by SIMS, electron holography, EBIC and 4D-STEM with precession diffraction.[14,19,22,32] The SIMS profile, provided as Supporting Information SI2, confirms *n* and *p* dopant concentrations $N_A = N_D = 9 \times 10^{18} cm^{-3}$ in the bulk areas, and shows that the junction does not present a sharp interface, but it is rather a *p-i-n* structure, with an intermediate region of about 200 nm.

Figure 2a shows the electric field maps measured by momentum-resolved 4D-STEM at different bias. Under forward bias, both the electric field and the depletion region width decrease, while under reverse bias, the electric field and the depletion region width increase. Figure 2b shows the variation of the electric field profile across the junction. A noise level of $0.03\ MV\,cm^{-1}$ was estimated by measuring the standard deviation (1σ) of the electric field maps in an area of 100 nm × 70 nm where the field was nominally zero. At zero bias, a depletion region width $W_d = (73 \pm 7)\ nm$ and a maximum electric field $E_{max} = (0.176 \pm 0.03)\ MV\,cm^{-1}$ were found. A table with the $E_{max}$ and $W_d$ extracted as a function of bias is shown in Supporting Information SI3. The built-in potential, $V_{bi}$, can be retrieved by integrating the electric field shown in Figure 2b. At zero bias, we estimate $V_{bi} = (0.88 \pm 0.06)\ V$. These results were compared to electron holography measurements performed in the same wafer (Supporting Information SI4), which lead to $V_{bi} = (0.86 \pm 0.05)\ V$ and $E_{max} = (0.175 \pm 0.02)\ MV\,cm^{-1}$,[14,19] in good agreement with 4D-STEM results. These values were obtained considering the 60 nm dead layer previously measured for the same wafer.[19]

In Supporting Information SI5, we recall the abrupt and linearly graded junction approximations. Both the 4D-STEM and electron holography results cannot be explained by the abrupt junction approximation, since the dopant concentration $N_A = N_D = 9 \times 10^{18} cm^{-3}$ obtained by SIMS (SI2) should lead to $W_d = 12\ nm$, $V_{bi} = 1.06\ V$ and $E_{max} = 0.859\ MV\,cm^{-1}$, as summarized in Table 1. Keeping to the abrupt model, the experimental results would rather correspond to a junction with a doping level in the range of $N_A = N_D = (2.8 - 4.5) \times 10^{17}\ cm^{-3}$ (see Table 1). The huge difference between nominal and experimental results can be explained by the presence of the intermediate region between the *p* and *n* layers. Looking at the SIMS profiles, a ≈ 73 nm wide depletion should be fully contained in the intermediate region, so that it could be modelled as a linearly graded junction. The experimental results can be reproduced assuming a linearly graded junction with a dopant variation per unit of length $\alpha = 1.5 \times 10^{23}\ cm^{-4}$ (see Table 1). However, using this model, the value of $V_{bi}$ remains on the higher part of the error bar of the experimental results. With this value of $\alpha$, and keeping in mind the doping levels in the bulk *n* and *p* regions ($N_A = N_D = 9 \times 10^{18} cm^{-3}$), the width of the linearly graded region should be $W_G = (N_A + N_D)/\alpha = 1.1\ \mu m$, which is inconsistent with SIMS measurements. Therefore, we assume that the junction is almost linearly graded in an intermediate region that extends over 200 nm, and the doping level increases rapidly outside this region.

As an additional information, the density of ionized dopants can be extracted from the derivative of the electric field shown in Figure 2a, using Poisson's equation. The result is shown in Figure 3a. The space charge region is clearly visible with a slight assymetry between *n* and *p* side, as observed in the SIMS data (Figure S2). In Figure 3b, the charge density profiles were averaged along the entire map, in order to reduce noise. At zero bias, we have found a maximum charge density $N_{Dmax} = (4.0 \pm 0.7) \times 10^{17}\ cm^{-3}$ on the *n* side and $N_{Amax} = -(3.1 \pm 0.7) \times 10^{17}\ cm^{-3}$ on the *p* side. These values are consistent with a built-in voltage $V_{bi} = 0.89\ V$, and confirm that the reduced electric field extracted form 4D-STEM is due to the small amount of active dopants, more than one order of magnitude smaller than the nominal value. This cannot be explained by partial ionization, looking at the low ionization energy of boron and phosphorous in silicon.[33] We also rule out a screening effect due to the electron beam, since our experiments were performed in the low injection regime, as shown in Supporting Information SI6. In summary, all the experimental results suggest that the reduced electric field and charge density, are indeed due to the graded nature of the junction.

We have verified that the linearly graded model is compatible with the variation of the electric field with bias. In a linearly graded junction, following eq. S12, $|E_{max}| \propto (V_{bi} - V)^{\frac{2}{3}}$. Figure 4 shows the variation of $|E_{max}|$ with bias compared with the theoretical values for different impurity grandients $\alpha$ in the range of $4 \times 10^{22}\ cm^{-4}$ to $4 \times 10^{23}\ cm^{-4}$. A good agreement is found for $\alpha = 1.5 \times 10^{23}\ cm^{-4}$, consistent with the above-described results.

The analytic equations in SI5 assume that the space charge region is fully depleted. As a consequence, the built-in potential is determined by the doping level at the edges of the depletion region ($x = \pm W_d/2$). Assuming $\alpha = 1.5 \times 10^{23}\ cm^{-4}$, we obtain $N(x = \pm W_d/2) = 5.9 \times 10^{17}\ cm^{-3}$, which leads to $V_{bi} = 0.918\ V$, as presented in Table 1. Both, this charge density and the built-in voltage, are higher than the estimations from 4D-STEM. To understand this difference, we have to keep in mind that the charge distribution at room temperature is more complex than the depletion approximation. A tail of electron and holes distribution penetrates the space charge region, whose limits are therefore not abrupt, and depend on the Debye length.[1] Using the charge density measured here, $N_{A,Dmax} \sim 3.0 - 4.0 \times 10^{17}\ cm^{-3}$, we can estimate $L_D \cong 7\ nm$, being the ultimate spatial resolution of the electric field maps presented here. In a linearly graded junction, this means that the maximum net charge density does not occur at the edges of the depletion region, but it is slightly shifted towards the junction interface. This is experimentally observed in Figure 3b, where the distance between the locations of maximum charge density is clearly smaller than $W_d$. As the charge density does not attain the maximum value of dopant concentration, located at ($x = \pm W_d/2$), this leads to a reduction of the built-in voltage.

To incorporate these thermal effects in our model, 1D simulations of the silicon *p-n* junction have been performed solving the Poisson equation numerically using the nextnano3 software. Figure 5a proposes a dopant profile (blue curve) that consists of a linearly graded junction with $\alpha = 1.5 \times 10^{23}\ cm^{-4}$, which extends over $W_G = 200\ nm$ (distance chosen to be coherent with the SIMS profile (SI2)). Outside this intermediate region, the dopant concentrations rapidly reach the bulk values: $N_A = N_D = 9 \times 10^{18}\ cm^{-3}$. Figures 5b and 5c present the comparison between simulated and measured electric field (absolute value) and charge density, respectively, at zero bias. The overall electrostatic parameters of the junction (electric field, depletion region and maximum charge density) are well reproduced with our extremely simplified model. A slightly better fit is obtained with $\alpha = 1.8 \times 10^{23}\ cm^{-4}$, instead of $\alpha = 1.5 \times 10^{23}\ cm^{-4}$. The fact that the two charge density maxima in Figure 5c are closer in the experimental curve than in the simulation is due to the fact that the real dopant profile is not perfectly linear, but sligtly superlinear.

It is worth highlighting that the 4D-STEM provides a direct measurement of the electric field (Figure 2b) and charge distribution (Figure 3b) in the space charge region, without any assumption in terms of doping profile, unlike other techniques like CV measurements[4,31] or EBIC[30]. With these experimental results and comparing with theoretical models, it is possible to determine the degree of ideality of the junction, detecting phenomena like dopant segregation or interdiffusion.

The main challenge for quantitatively evaluating the electric field by 4D-STEM is the unknown value of the electrically inactive sample thickness[13], which is often speculated to be at the origin of measuring a lower built-in potential than expected in electron holography.[13] However, the error associated with the surface dead layer can be identified and corrected by measuring the electron beam deflection or phase shift as a function of the sample thickness, as demonstrated both for 4D-STEM[26] and electron holography.[13,14,34] Here, we have demonstrated that the measurement of an electric field lower than the nominal value is not necessarily an artifact: it can originate from a non-ideal doping profile. Careful 4D-STEM measurements incorporating the dead layer thickness correction and in-situ biasing can provide valuable information about the doping profile.

In summary, we have combined in-situ biasing and momentum-resolved 4D-STEM measurements of a symmetric silicon p-n junction, previously analyzed by SIMS and electron holography. We have obtained a maximum electric field $E_{max} = (0.176 \pm 0.03)\ MV cm^{-1}$, in agreement with electron holography results. The charge density distribution in the space charge region was directly extracted from the 4D-STEM measurement of the electric field. The comparison of the measured values with analytical equations, as well as finite element simulations, demonstrate that the dopant density profiles do not drop abruptly at the junction. Instead, this sample can be reasonably well modelled as a linearly graded p-n junction with an impurity gradient in the range of $\alpha = (1.5 \pm 0.7) \times 10^{23}\ cm^{-4}$.

The development of more efficient semiconductor devices, currently reaching the nanoscale, demands tools able to provide fast feedback for further device optimization. Our results show that in-situ biased 4D-STEM is a suitable approach to study the interface quality of p-n junctions, enabling the measurement of crucial electrical properties of such devices.

**Supplementary Material**

See supplementary material for additional information of the measurements and details of the calculations presented here.

**Acknowledgements**

Martien I. den-Hertog, B. C. da S., Z. S. M., H. O., E. M. and J. L. R. thank the European Union's Horizon 2020 research and innovation programme under the European Research Council (ERC) Grant agreement N° 758385 (e-See). These experiments have been performed at the Nanocharacterisation platform (PFNC) based at Minatec, Grenoble.

**Data Availability Statement**

The data that support the findings of this study are available from the corresponding authors upon reasonable request.

**Notes**

The authors declare no competing financial interest.

# TABLES

**Table 1** – Theoretical calculations of the maximum electric field, depletion region width and built-in potential for a silicon p-n junction with different doping levels, considering the abrupt and linearly graded approximations. The equations derived for both approximations can be found in the Supplementary Information (SI5). The experimental results have been added for comparison.

|  | $E_{max}$ ($MVcm^{-1}$) | $W_d$ (nm) | $V_{bi}$ (V) |
|---|---|---|---|
| Experimental (4D-STEM) | 0.176 ± 0.03 | 73 ± 7 | 0.88 ± 0.06 |
| Experimental (Electron Holography) | 0.175 ± 0.02 | -- | 0.86 ± 0.05 |
| *Abrupt p-n junction* ($N_A = N_D = 9.0 \times 10^{18} cm^{-3}$) | 0.859 | 12 | 1.06 |
| *Abrupt p-n junction* ($N_A = N_D = 4.5 \times 10^{17} cm^{-3}$) | 0.178 | 51 | 0.906 |
| *Abrupt p-n junction* ($N_A = N_D = 2.8 \times 10^{17} cm^{-3}$) | 0.138 | 64 | 0.881 |
| *Linearly graded p-n junction* ($\alpha = 1.5 \times 10^{23} cm^{-4}$) | 0.176 | 78 | 0.918 |

**FIGURE CAPTIONS**

**Figure 1** – Schematic of the momentum-resolved 4D-STEM experiment performed in a silicon *p-n* junction. Reverse bias is obtained by applying a negative bias ($-V_{bias}$) to the *p*-side while the *n*-side is grounded.

**Figure 2** – (a) In-situ biased 4D-STEM electric field maps of the silicon *p-n* junction. (b) Profiles of the electric field obtained from the maps shown in (a) by integration along the entire map, as indicated in (a). The measured depletion length for zero bias is indicated, following the method used by.[16]

**Figure 3** – (a) In-situ biased 4D-STEM charge density maps of the silicon *p-n* junction. (b) Profiles of the charge density obtained from (a) by integration along the entire map, as indicated in (a).

**Figure 4** – Comparison between the values of maximum electric field measured by in-situ biased 4D-STEM and the expected evolution with bias considering a linearly graded p-n junction with various impurity gradients, *α*.

**Figure 5** – Dopant profile (blue), $N_D - N_A$, with an intermediate graded region extending over 200 nm consisting of an linearly graded p-n junction with an impurity gradient of $\alpha = 1.5 \times 10^{23} cm^{-4}$. Outside the intermediate region, the dopant concentration is the bulk value $N_A = N_D = 9 \times 10^{18} cm^{-3}$. (b) and (c) Comparison between simulated and experimental electric field and charge density at zero bias, respectively. In (b) and (c), we demonstrate that an impurity gradient of $\alpha = 1.8 \times 10^{23} cm^{-4}$ (magenta) provides a better fit due to the partial compensation of thermal effects.

**FIGURE 1**

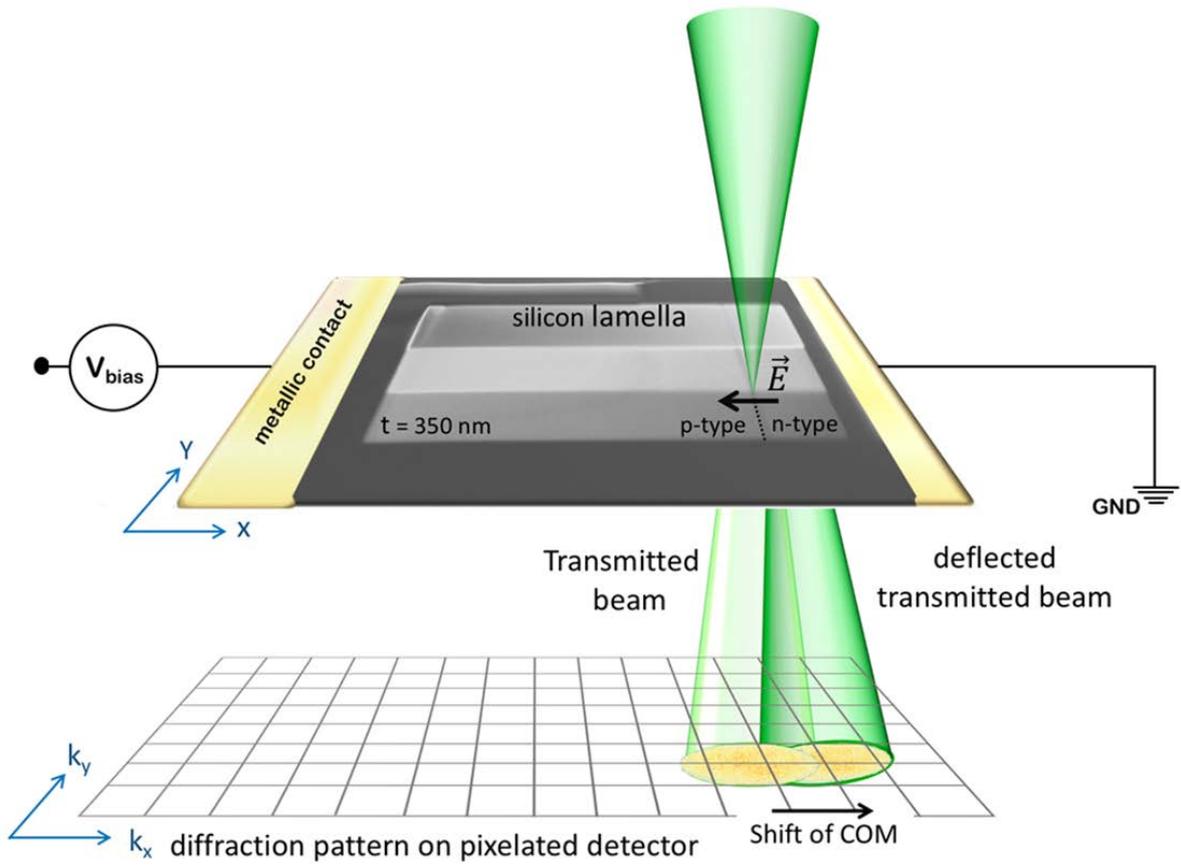

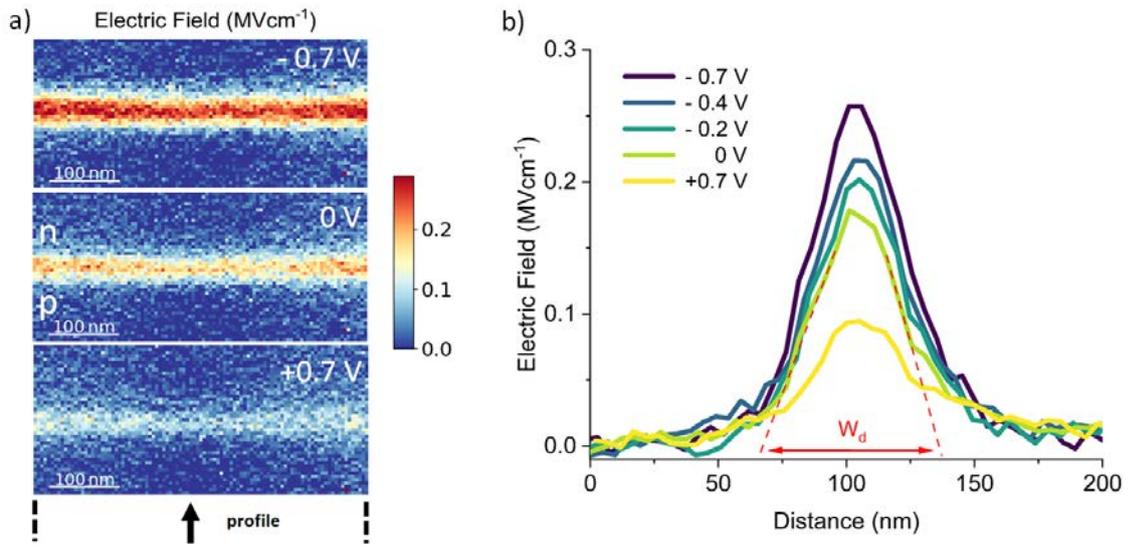

**FIGURE 3**

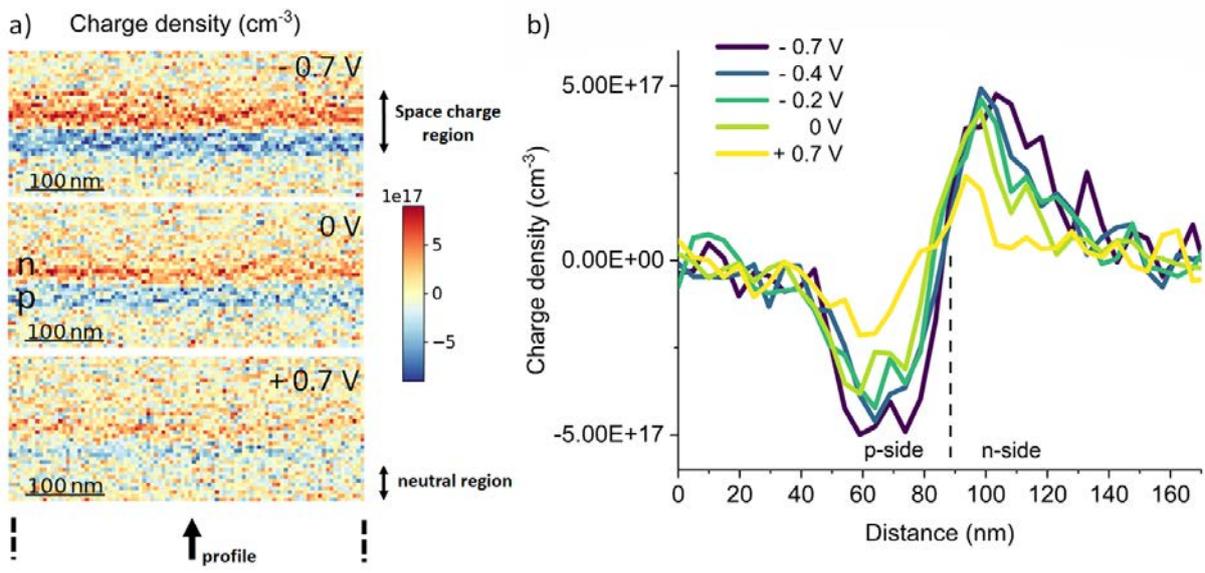

**FIGURE 4**

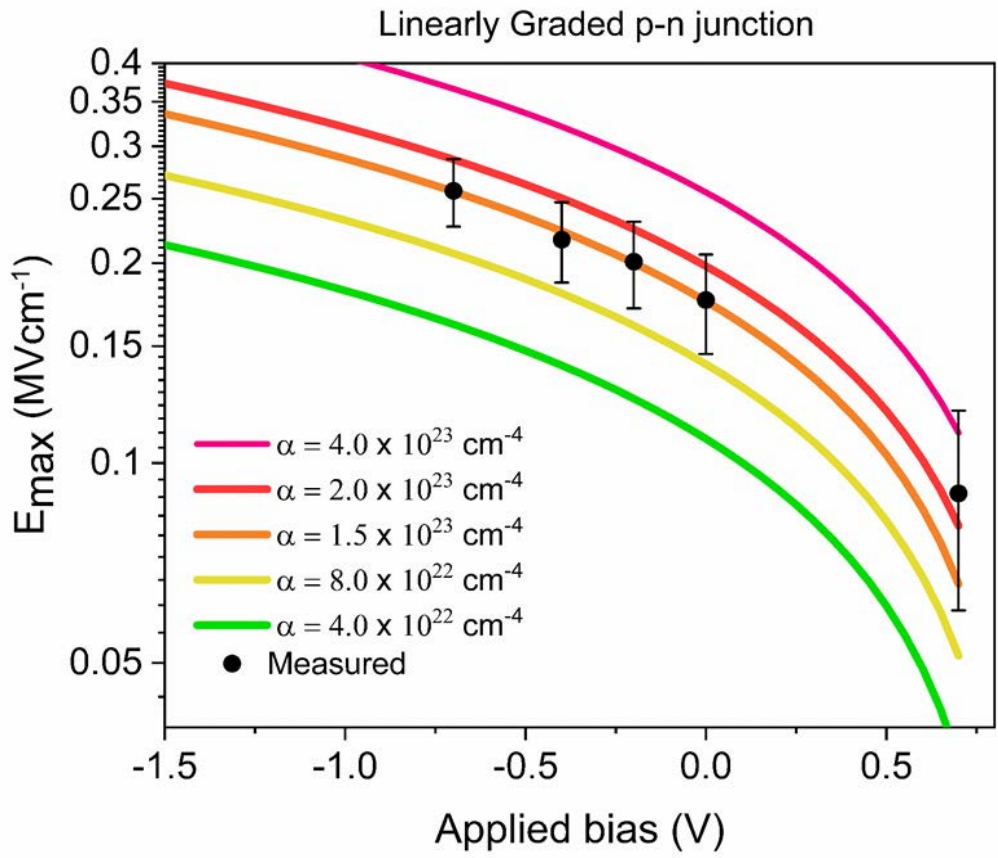

**FIGURE 5**

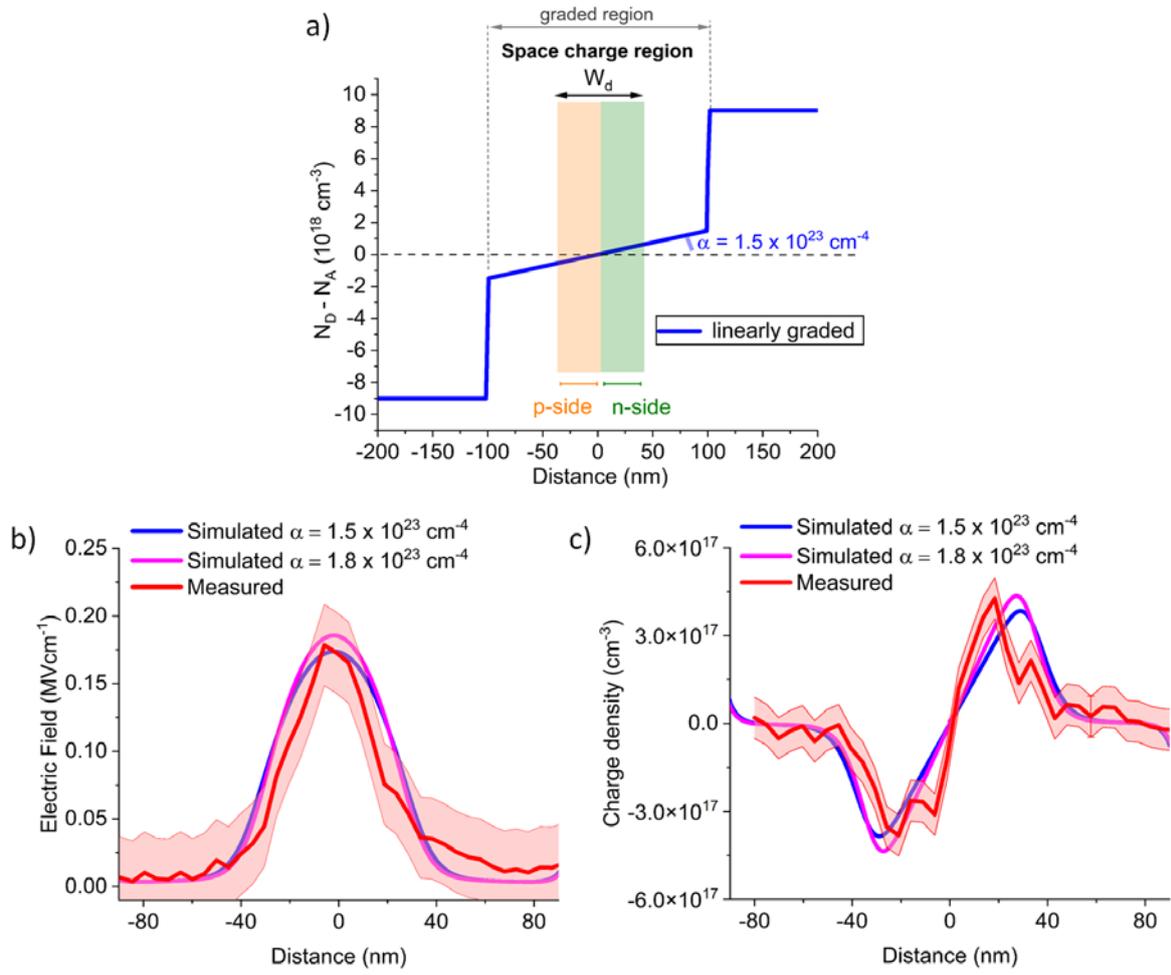